\documentclass{elsarticle}

\usepackage{color}
\usepackage{hyperref}

\def\cm{cm$^{-1}$}
\def\sb{SmB$_6$}
\def\sba{Al Flux-SmB$_6$}
\def\sbp{FZ SmB$_6$-Pure}
\def\sbd{FZ SmB$_6$-Def}

\begin{document}

\begin{frontmatter}

\title{An effect of Sm vacancies on the hybridization gap in topological Kondo insulator candidate \sb}

\author[mymainaddress]{Michael E. Valentine}

\author[mymainaddress]{Seyed Koohpayeh}

\author[mymainaddress]{W. Adam Phelan}

\author[mymainaddress]{Tyrel M. McQueen}

\author[mysecondaryaddress,mysecondaryaddress1]{Priscila F. S. Rosa}

\author[mysecondaryaddress]{Zachary Fisk}

\author[mymainaddress]{Natalia Drichko*}
\cortext[mycorrespondingauthor]{Natalia Drichko}
\ead{drichko@jhu.edu}

\address[mymainaddress]{Institute for Quantum Matter and Department of Physics and Astronomy, Johns Hopkins University, Baltimore, MD
21218, USA}
\address[mysecondaryaddress]{Department of Physics and Astronomy, University of California, Irvine, California 92697, USA}
\address[mysecondaryaddress1]{Los Alamos National Laboratory, Los Alamos, NM 87545, USA}

\begin{abstract}

A necessary element for the predicted topological state in Kondo insulator \sb\ is the hybridization gap which opens in this compound at low temperatures. In this work, we present a comparative study of the in-gap density of states due to Sm vacancies by Raman scattering spectroscopy and heat capacity for samples where the number of Sm vacancies is equal to or below 1~\%. We demonstrate that hybridization gap is very sensitive to the presence of Sm vacancies. At the amount of vacancies above 1~\%\ the gap fills in with impurity states and low temperature heat capacity is enhanced.
\end{abstract}


\end{frontmatter}


\section{Introduction}

The theoretical and experimental discovery of topological insulators is one of the most exciting recent achievements in the field of condensed matter physics. Originally, topological surface states were found in  insulators with strong spin-orbit coupling where an electronic gap with band inversion exists in band structures. Later, it was suggested that materials where a gap with band inversion results from strong electron-electron interactions, so called Kondo insulators, can demonstrate topological surface states~\cite{Dzero2010,Dzero2012,Alexandrov2013}.  One example of such a system   is the Kondo insulator \sb, where a gap opens at the Fermi level below 70~K due to hybridization of the $4f$ and $5d$ electronic bands. Apart from resistivity measurements, which detect the activated behavior below 70 K, the hybridization gap was detected by various   methods \cite{Travaglini1984,Ohta1991,Nanba1993,Gorshunov1999,Jiang2013,Flachbart2001,Frankowski1982,Wolgast2013,Valentine2016}\textbf{.}

\sb\ crystals grown through the floating zone method tend to  possess some amount of vacancies at the Sm sites~\cite{Phelan2015}. With a presence of Sm vacancies the resistivity temperature dependence preserves its activated behavior \cite{Phelan2015}, while a difference is observed in the shape of the resistivity plateau at low temperatures depending on the number of Sm vacancies. According to Ref.~\cite{Phelan2015} the plateau  appears at lower resistivity values and is less pronounced for samples with a higher concentration of Sm vacancies. Our results  presented in this work demonstrate that the difference is even more striking when the low temperature electronic structure and in-gap density of states (DOS) are probed by Raman spectroscopy and heat capacity.

An understanding of the dependence of the hybridization gap and in-gap density of states on the presence of Sm vacancies is important for identification of topological Kondo insulating states. It is also necessary for determining the source of the metallic surface states~\cite{Zhu2013,Syers2015}, which can result from topologically trivial effects,  and for verifying the cause of such unexpected effects as 3D quantum oscillations observed in \sb~\cite{Tan2015}.

Raman scattering allows to simultaneously probe Sm vacancies through phonon effects, and electronic structure through electronic Raman response. This gives an advantage over such methods of sample characterization as resistivity, which do not give any insight into microscopic properties of the samples. In Ref.~\cite{Valentine2016}, we demonstrated a way to characterize the number of Sm vacancies in \sb\ by the intensity of the Raman forbidden phonon at about 10~meV associated with off $\Gamma$-point scattering of acoustic phonons~\cite{Alekseev2015} which arises in samples with Sm vacancies due to local symmetry breaking.  In this work, we performed a comparative study of three \sb\ samples: the \sba\ sample contains negligible number of vacancies, the \sbd\ sample contains  about 1~\%\ of Sm vacancies, and the \sbp\   about 0.5~\%\ of Sm vacancies. We show that our Raman results on in-gap density of states are in agreement with our heat capacity results.

\section{Experimental}

\sb\ samples were synthesized by floating zone (\sbp\ and \sbd) and Al flux (\sba) methods. For \sbd\textbf{,} the number of Sm vacancies of about 1~\%\ was estimated by measuring the size of lattice constant \cite{Phelan2015}.  For all samples, Raman spectroscopy was used to approximate the number of vacancies according to Ref.~\cite{Valentine2016}.

Raman scattering spectra were measured using Horiba Jobin-Yvon T64000 triple monochromator spectrometer  using  the 514 nm line of  a Coherent Ar$^+$ laser as an excitation. A ST-500 Janis cold finger cryostat was used to cool down the sample to 4~K. For more details on the measurements see~\cite{Valentine2016}. All the data presented in this paper were measured in $A_{1g}+E_g$ symmetry on single crystals of \sb\ oriented by X-ray and polarization dependent Raman scattering measurements.

The heat capacity of the \sb\ samples used for Raman scattering experiments was measured in the temperature range between 80 and 2~K using Quantum Design PPMS.

\section{Results}

Raman spectra of \sb\ show phonon bands superimposed on electronic background. The three intense bands observed above 700~\cm\ are due to  scattering from phonons which involve motion of the B$_6$ octahedra~\cite{Valentine2016}. The lower frequency spectra discussed in this paper are dominated by Raman-forbidden and two-phonon scattering from acoustic phonons, the spin exciton observed below approximately 20~K, and an electronic background resulting from interband transitions.

\begin{figure}
  \centering
  \includegraphics[width=12cm]{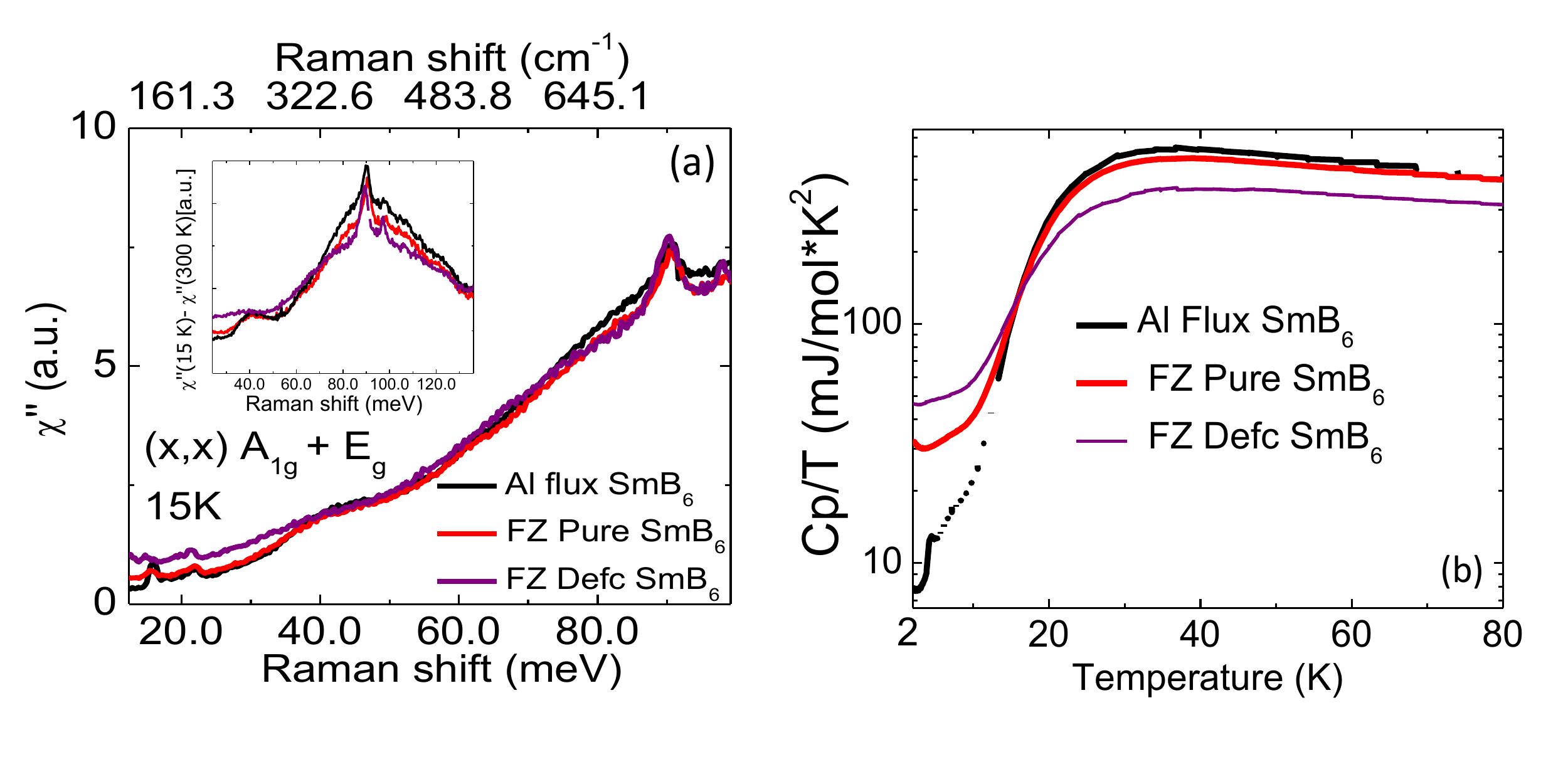}\\
  \caption{ (a) Raman scattering spectra for the \sba, \sbp, and \sbd\ samples at 15~K in the spectral range relevant to the hybridization gap. The inset shows the same data with the room temperature electronic background extracted. Bands at 40 and 100 meV corresponding to Raman scattering across the hybridization gap are clearly observed in \sba\ sample. Raman intensity in the gap increases and 40~meV transition gets smeared with the increase of the number of vacancies. (b) Temperature dependence of the heat capacity of \sba, \sbp, and \sbd\ samples.}\label{HC}
\end{figure}

As the samples are cooled below 150~K, the electronic Raman response shows an opening of hybridization gap  as a depression in the electronic background at low frequencies, and a formation of two broad bands at 100 and 40~meV which (Fig.~1(a)) correspond to the Raman scattering from the two hybridized bands  below E$_F$ to the one above E$_F$ as suggested by  the band structure calculations of Ref.~\cite{Lu2013}. The maximum at 40 meV corresponds to the transitions across the hybridization gap. We show these bands with room temperature background extracted in Fig.1 (a) (inset).

\begin{figure}
  \centering
  \includegraphics[width=12cm]{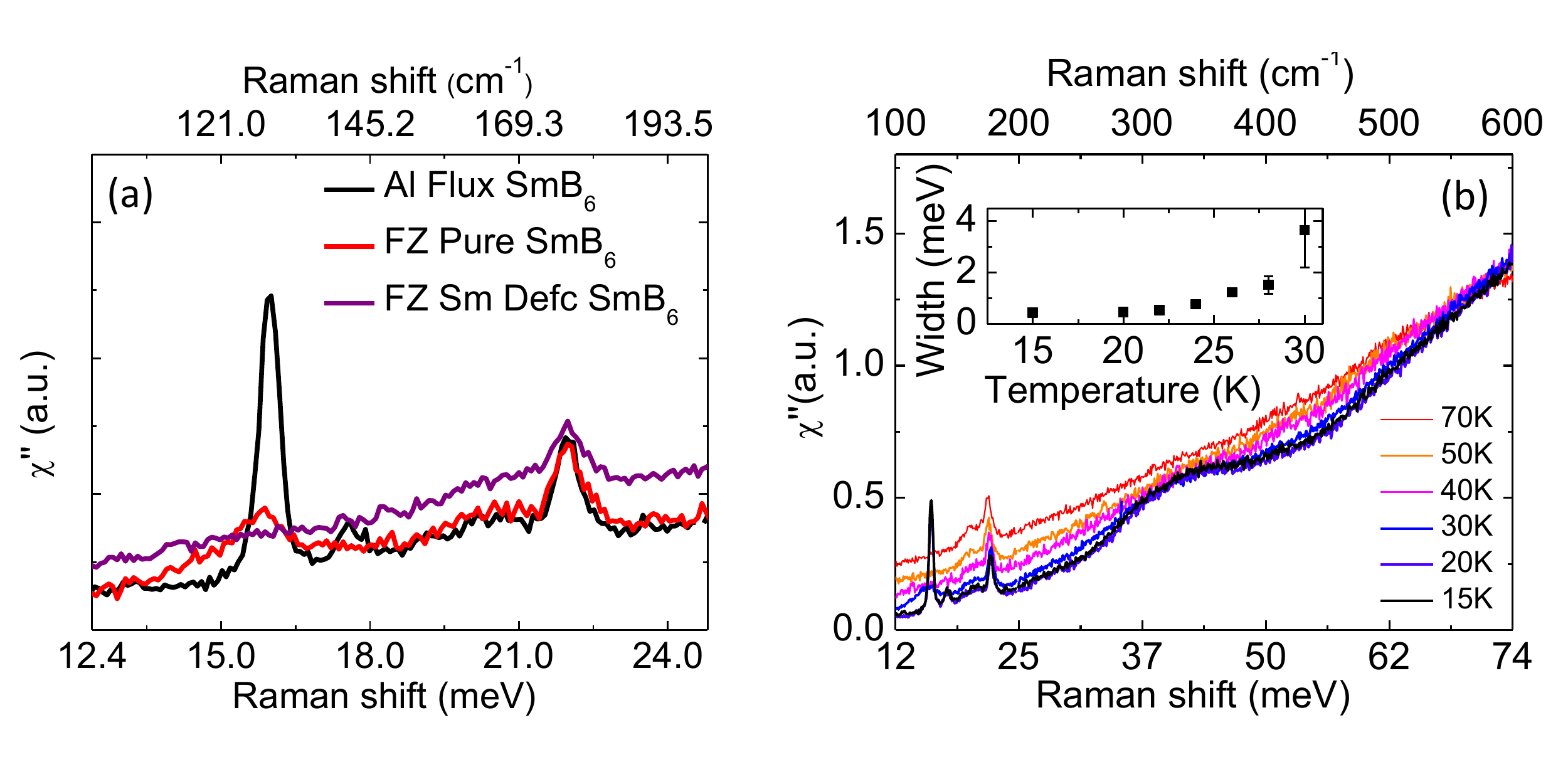}\\
  \caption{(a) Raman scattering of the \sba, \sbp, and \sbd\ samples at 15~K in  the spectral range relevant to spin exciton. Note the increase of the width and quenching of the exciton with increasing number of Sm vacancies. (b) Temperature dependence of the exciton band for the \sba\ sample. The inset shows the temperature dependence of the width of the exciton band. }\label{HC}
\end{figure}

As seen in Fig.~1, the hybridization gap below 20~meV fully opens only in the \sba\ samples, where the number of Sm vacancies is very small and comparable to the noise of our measurements. As the number of vacancies increases up to approximately 0.5~\%\ in the \sbp\ sample, we observe more Raman intensity below 20~meV, and the gap feature at 40~meV is smeared. In the \sbd\ sample, the feature at 40~meV is not distinguishable, and even higher intensity in the range below 20~meV is observed. For the \sba\ sample, at low temperatures we observe a very narrow feature at 16~meV (see Fig.~2). No phonon features are expected in this range, while   neutron scattering~\cite{Fuhrman2015} and theory predictions~\cite{Fuhrman2014} suggest a presence of a spin exciton.

We also observe a strong effect of the presence of Sm vacancies on the spin exciton feature (Fig.~2). It appears in the spectra of the \sba\ sample as a very narrow (0.5~meV) band at about 16~meV. It is widened up to 1.5~meV for \sbp\ sample and is absent from the spectra of \sbd.

This striking difference between these three samples is also observed in the heat capacity data shown in Fig~1(b). The overall shape of the temperature dependence is the same, with C$_p$ going through a maximum at around 40~K associated with the opening of hybridization gap, decreasing at temperatures around 20~K, and flattening again at about 5~K in the low temperature metallic regime. A small broadening of the maximum of heat capacity at around 40~K is observed from \sba\ to \sbp\ sample, while this maximum becomes considerably broader with lower $C_p$ values for \sbd. At around 2~K $C_p/T$ of the  \sba\ sample reaches the lowest value of 7 mJ/mol*K$^2$ observed for \sb\ samples~\cite{Gabani2002}. The low temperature values of C$_p/T$ observed for \sbp\ are about 30 mJ/mol*K$^2$, while the highest values of C$_p/T$ of about 45 mJ/mol*K$^2$ are observed for the \sbd\ sample.  A spread of the low temperature heat capacity values was previously observed in literature~\cite{Gabani2002}. Our measurements in which we use samples with a calibrated number of Sm vacancies demonstrate that this spread originates from the low frequency density of states due to presence of Sm vacancies.

\section{Discussion}

Both Raman scattering data and heat capacity of the measured \sb\ samples show a striking change of the low temperature  behavior with an increase of the number of Sm vacancies. Raman scattering clearly shows an appearance of electronic states below the frequencies of the hybridization gap (20 meV) with an increase of the number of Sm vacancies.  The same information is revealed by the increase of $C_p$ values  with an increase of Sm vacancies in the low temperature regime. A theoretical insight on the nature of electronic states in hybridization gap  due to non-$f$-electron impurities in Kondo insulating systems~\cite{Riseborough2003} suggests a growth of an impurity band as a function of the increasing impurities concentration.

Heat capacity consists of electron and phonon contributions of the form $C_p = C_p(e)+C_p(ph)$. There is no reason for the phonon contribution $C_p(ph)$ for these three samples to be different, thus we can focus on the electronic contribution.  At temperatures above 2~K the main contributions to electronic part of heat capacity $C_p(e)=\gamma T + exp(-\Delta/T)$ are the free electron contribution and a term associated with the opening of the hybridization gap $\Delta$, which  appears in the temperature dependence as a maximum at around 40~K.  The broadening of this maximum in $C_p$ with the increase of the number of Sm vacancies indicates that the opening of the hybridization gap is suppressed. Indeed, Raman scattering  reveals that at 1~\%\ of Sm vacancies the maximum in Raman electronic response   at 40 meV corresponding to excitations over hybridization gap becomes indistinguishable, and Raman intensity appears at in-gap frequencies, indicating electronic states at the hybridization gap energies. A change of the density of states which corresponds to this change of the Raman scattering electronic response, an increase of the in-gap DOS and a simultaneous decrease of DOS at the gap edges are predicted in Ref.~\cite{Riseborough2003} as the result of the doping.

It is interesting to compare our results to resistivity presented for \sbp\ and \sbd\ samples in \cite{Phelan2015}. According to \cite{Phelan2015}, \sbp\ and \sbd\ samples show a similar behavior of resistivity, with increase by few orders of magnitude at about 40 K. The difference between the samples is observed in resistivity values at 10 K and lower. This is in contrast to our Raman scattering and heat capacity data, where the difference between the samples in clearly observed already a 40 K.

The heat capacity below 10~K can be roughly described by the $\gamma T$ contribution of free electrons, due to the impurity band which crosses $E_f$ in hybridization gap. Here we neglect  the fact that in \sb\ an increase of $C_p$ is found below 2~K~\cite{Phelan2014,Gabani2002}. The values of $\gamma$ for the three measured samples are presented in Table~\ref{gammas}. The smallest value is observed for \sba\, while for the other two samples $\gamma$ increases reflecting the large increase of the electronic density of states    of metallic nature.
The presence of electronic states in the hybridization gap are also evidenced both by the larger intensity of Raman scattering below 20~meV for \sbp\ and \sbd\ samples compared to \sba.

\begin{table}
\caption{Values of coefficient $\gamma$ for $C_p$ and with of the in-gap exciton for the three studied \sb\ samples.}
\begin{tabular}{|c|c|c|c|}
  \hline
               &                    \sba & \sbp & \sbd \\ \hline
$\gamma$ ($C_p$)                    & 2 mJ mol$^{-1}$ K$^{-2}$ & 24 mJ mol$^{-1}$ K$^{-2}$ & 42 mJ mol$^{-1}$ K$^{-2}$ \\ \hline
$\Gamma$ (exciton line width)       & 0.5 meV & 1.5 meV & - \\ \hline
\end{tabular}
\label{gammas}
\end{table}

Another measure of the DOS inside the hybridization gap is the width of the band of the exciton observed at around 140~\cm\ (16 meV) (Fig.~2). This band   is observed as an exceptionally narrow feature in the spectra of \sba\ at low temperatures. It is assigned to the scattering on the exciton level which is formed inside the hybridization gap due to electron-electron repulsion~\cite{Valentine2016,Nyhus1997,Fuhrman2015}. The  small width $\Gamma$ of the band is defined by the long life time of the excitation $\tau$=$1/\Gamma$. The relaxation time of such a level would depend on the probability of non-radiative decay as $1/\tau = \Gamma \sim 4 \pi (J_{ex}N(0))^2T$~\cite{Becker1977}. Here $N(0)$ is the density of states of the impurity band in hybridization gap, to which the exciton can couple,   and  $J_{ex}$ a coupling constant which allows the decay.   The increase of $\Gamma$ by three times from \sba\ to \sbp\ sample indicates the respective increase of DOS $N(0)$(\sbp)/$N(0)$(\sba) = 1.5. While the tendency of the increase of the in-gap DOS is reproduced, the value suggested by this simple calculation is much smaller than that   suggested by $\gamma$ values of  $C_p$.  One reason for it could be that the width of the exciton band $\Gamma$ for \sba\ sample at the lowest temperature is defined by the integration over some part of BZ, and not by the natural width. The exciton band for a very pure \sb\ sample measured by neutron scattering which provides the natural width due to   $q$ resolution is below 0.25 meV~\cite{Fuhrman2015}.

This point is supported by the temperature dependence of the exciton band width followed for the \sba\ sample (Fig.~2(b)). $\Gamma$ (T) decreases faster on cooling than linearly with temperature as  suggested   by the formula that defines the decay process of the exciton band.  One possibility is that this dependence is defined by the decrease of the in-gap N(0) on the opening of the hybridization gap. Flattening of this dependence below 20~K indicates the temperature below which the natural width of the band becomes smaller than the measured one defined by the dispersion of the exciton level.

\section{Conclusions}

In this paper, we demonstrated an increase of the density of states inside the hybridization gap  with an increase of Sm vacancies in \sb\  using Raman scattering spectroscopy and heat capacity.   The density of states inside the hybridization gap increases with an increase of the number of vacancies, and fills in the gap when the concentration of Sm vacancies reaches  1~\%.
We compare a quantitative estimate of the $N(0)$ increase received from Raman scattering and heat capacity.

Our results are in agreement with the qualitative picture of the growth of the impurity band inside the hybridization gap on the doping of Kondo insulator with non-$f$ electron impurities provided by theoretical work in Ref.~\cite{Riseborough2003}. However, theory  suggests the  closing of  hybridization gap   should occur at a  concentration of vacancies at or above 7~\%, while we observe it at much lower values of 1~\%. At this point it remains  a question  whether the fact that Sm$^{+3}$ possesses magnetic moment while Sm$^{+2}$ does not is of an importance to explain this discrepancy.

\section{Acknowledgements}

We are grateful to C. Broholm, P. Nikoli\'c, W. Fuhrman,  J. Paglione, and N. P. Armitage for useful discussions. The work
at IQM was supported by the U.S.
Department of Energy, Office of Basic Energy Sciences, Division of Material Sciences and Engineering under Grant No.
DE-FG02-08ER46544. The work at Los Alamos National Laboratory was performed under the auspices of the U.S. DOE through the Los Alamos LDRD program​.


\bibliography{SmB6}

\end{document}